\documentclass[aps,prl,twocolumn,superscriptaddress]{revtex4}

\usepackage{graphicx}

\begin{document}

\preprint{}

\title{Enhancement of Superconducting Transition Temperature due to the strong Antiferromagnetic Spin Fluctuations in Non-centrosymmetric Heavy-fermion Superconductor CeIrSi$_3$ :A $^{29}$Si-NMR Study under Pressure}

\author{H. Mukuda}
\email[]{e-mail  address: mukuda@mp.es.osaka-u.ac.jp}
\author{T. Fujii}
\author{T. Ohara}
\author{A. Harada}
\author{M. Yashima}
\author{Y. Kitaoka}
\affiliation{Department of Materials Engineering Science, Graduate School of Engineering Science, Osaka University, Osaka 560-8531, Japan}

\author{\\Y. Okuda}
\author{R. Settai}
\author{Y. Onuki}
\affiliation{Department of Physics, Graduate School of Science, Osaka University, Osaka 560-8531, Japan}

\date{\today}

\begin{abstract}
We report a $^{29}$Si-NMR study on the pressure-induced superconductivity (SC) in an antiferromagnetic (AFM) heavy-fermion compound CeIrSi$_3$ without inversion symmetry. In the SC state at $P=$2.7-2.8 GPa, the temperature ($T$) dependence of the nuclear-spin lattice relaxation rate $1/T_1$ below $T_{\rm c}$ exhibits a $T^3$ behavior without any coherence peak just below $T_{\rm c}$, revealing the presence of line nodes in the SC gap. In the normal state, $1/T_1$ follows a $\sqrt{T}$-like behavior, suggesting that the SC emerges under the non-Fermi liquid state dominated by AFM spin fluctuations enhanced around quantum critical point (QCP). The reason why the maximum $T_{\rm c}$ in CeIrSi$_3$ is relatively high among the Ce-based heavy-fermion superconductors may be the existence of the strong AFM spin fluctuations. We discuss the comparison with the other Ce-based heavy-fermion superconductors.
\end{abstract}


\maketitle

\begin{figure}[htbp]
\centering
\includegraphics[width=70mm]{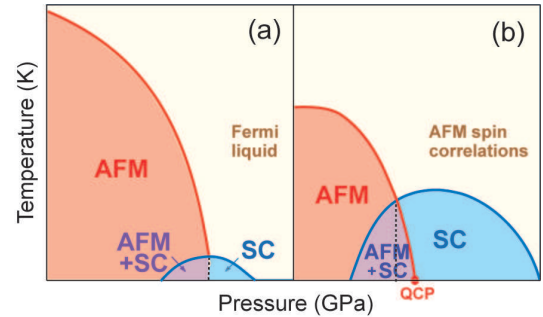}
\caption[]{
(Color online) Schematic phase diagrams of HF compounds\cite{Kitaoka1}. (a) The $P \,$-$\, T$ phase diagram for CePd$_2$Si$_2$, CeIn$_3$ and CeRh$_2$Si$_2$, and (b) for CeCu$_2$Si$_2$ and CeRhIn$_5$. The SC in the group-(a) emerges under the Fermi-liquid state without any trace of AFM spin fluctuations, while the SC in the group-(b) does under the non-Fermi liquid state dominated by AFM spin fluctuations. The $T_{\rm c}$ of the group-(b) is known to be higher than that for that of the group-(a).
}
\label{Yashima-fig1}
\end{figure}

A number of studies on heavy-fermion (HF) compounds revealed that unconventional superconductivity (SC) arises at or close to a quantum critical point (QCP), where magnetic order disappears at low temperature as a function of pressure ($P$) \cite{Jaccard92,Movshovic96,Grosche,Mathur98,Hegger00}. These findings suggest that the mechanism forming Cooper pairs can be magnetic in origin. 
However, the nature of SC and magnetism is still unclear when SC appears very close to the antiferromagnetism (AFM). 
The phase diagram, schematically shown in Fig.~\ref{Yashima-fig1}(a) \cite{Kitaoka1}, has been observed in HF AFM compounds such as CeIn$_3$ \cite{Mathur98}, CePd$_2$Si$_2$ \cite{Grosche} and CeRh$_2$Si$_2$ \cite{Movshovic96,Araki}. In CeIn$_3$ \cite{SKawasaki1} and CeRh$_2$Si$_2$ \cite{Araki}, the $P$-induced first-order transition from AFM to paramagnetism have been revealed near the boundary where SC emerges without the development of AFM spin fluctuations. Remarkably different behavior, schematically shown in Fig.~\ref{Yashima-fig1}(b), has been found in the archetypal HF superconductors CeCu$_2$Si$_2$ \cite{Steglich,YKawasaki1,YKawasaki2} and CeRhIn$_5$ \cite{Hegger00,Muramatsu1,YashimaPRB2007}. Although an analogous behavior relevant with an AFM-QCP has been demonstrated in both the compounds, it is noteworthy that the associated SC region extends to higher pressures than in the other compounds, their value of $T_c$ reaching its maximum away from the verge of AFM\cite{Muramatsu1,YashimaPRB2007,Bellarbi,Thomas}. 

\begin{figure}[htbp]
\begin{center}
\includegraphics[width=0.9\linewidth]{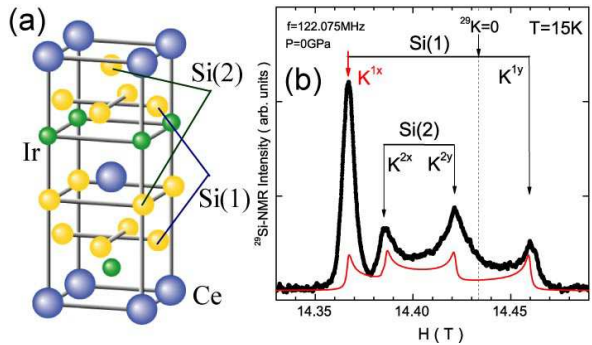}
\end{center}
\caption[]{(Color online) (a) Crystal structure of CeIrSi$_3$. There are two inequivalent Si sites denoted as Si(1) and Si(2). (b) The Si-NMR spectrum in the paramagnetic state at $P=$ 0. The solid curve is a simulation of two dimensional powder pattern of the NMR spectrum for Si(1) and Si(2) where the powder is preferentially oriented with the magnetic field in the basal plane (see the text). }
\label{fig:spectra}
\end{figure}

Recently, $P$-induced SC was discovered in the HF AFM compounds CeRhSi$_3$ \cite{Kimura} and CeIrSi$_3$ \cite{Sugitani}. Remarkably, the SC transition temperature $T_{\rm c}$= 1.6 K for CeIrSi$_3$ under $P$= 2.6 GPa is relatively high among the Ce-based HF superconductors. Notably, as shown in Fig.\ref{fig:spectra}(a), the inversion symmetry along the c-axis is broken in these compounds with a tetragonal structure as well as in CePt$_3$Si\cite{Bauer}.  
In superconductors that lack an inversion symmetry, the relationship between a spatial symmetry and the Cooper pair spin state may be broken, which makes the parity of SC state mixed between even and odd parities through the Rashba-type antisymmetric spin-orbit coupling (ASOC) \cite{Gorkov,Frigeri,Samokhin} and hence leads to a mixed order parameter of spin-singlet and spin-triplet Cooper pairing states. The discovery of SC in the non-centrosymmetric HF compound CePt$_3$Si \cite{Bauer} has attracted considerable attention both theoretically \cite{Frigeri,Samokhin,Hayashi,Fujimoto,Yanase} and experimentally \cite{Yogi2004,Bonalde,Izawa} from this point of view. The intrinsic SC characteristics of CePt$_3$Si, however, are still controversial \cite{Takeuchi}. In this context, a new state of SC would be expected for the $P$-induced HF superconductors CeRhSi$_3$ and CeIrSi$_3$ without inversion symmetry.


A polycrystalline sample of CeIrSi$_3$ was grown using a tetra-arc furnace, as described elsewhere \cite{Sugitani} and moderately crushed into large size grains in order to make RF pulses for NMR measurements penetrate the samples easily. Hydrostatic pressure was applied by utilizing a NiCrAl-BeCu piston-cylinder cell filled with an Si-based organic liquid as the pressure-transmitting medium. To calibrate the pressure at low temperatures, the shift in $T_{\rm c}$ of Sn metal under $P$ was monitored by its resistivity.

\begin{figure}[htbp]
\begin{center}
\includegraphics[width=0.9\linewidth]{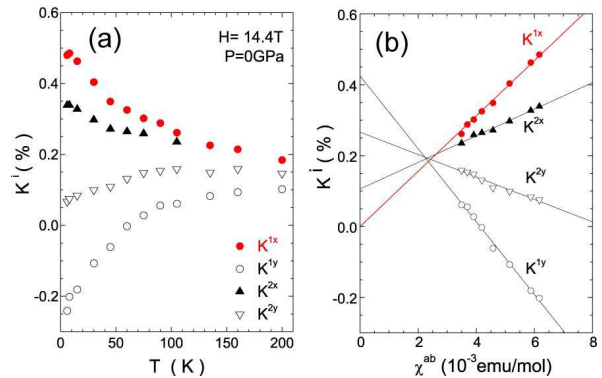}
\end{center}
\caption[]{(Color online) (a) $T$ dependence of Knight shift for the Si(1) and Si(2) sites. (b) $K^i$s ($i=1x$, $1y$, $2x$, and $2y$) are plotted against the bulk susceptibility in the basal plane ($\chi^{\rm ab}$) with temperature as an implicit parameter. The $T$-independent orbital shifts are estimated from the intercepts of these plots (see the text). }
\label{fig:K-chi}
\end{figure}

Figure \ref{fig:spectra}(b) shows a typical field-swept $^{29}$Si-NMR spectrum in the paramagnetic state at ambient pressure ($P=0$). Two sets of NMR spectra are observed with two peaks originating from two inequivalent Si sites denoted as Si(1) and Si(2). Each NMR spectral shape with two peaks is inferred to be a basal-plane oriented powder pattern. Here the basal plane is preferentially oriented by the magnetic field, since the susceptibility $\chi^{\rm ab}$ in the basal plane is significantly larger than  $\chi^{\rm c}$ along the c-axis \cite{Sugitani}.  Noting that the number of Si(1) sites is twice that of the Si(2) sites, the NMR spectra originating from these sites can be differentiated as indicated in Fig. \ref{fig:spectra}(b). In fact, these spectra are simulated by assuming the basal-plane oriented powder pattern, as shown by the solid curve in the figure. The two sharp peaks for the Si(1) and Si(2) sites originate from the grains with a field parallel to the two principal axes of the Knight-shift tensor in the basal plane. Hereafter the positive and negative components of the Knight shift are denoted as $K^{1x}$ and $K^{1y}$ ($K^{2x}$ and $K^{2y}$) with the fields parallel to the $x$- and $y$-axis for the Si(1)(Si(2)) site, respectively. The temperature ($T$) dependence of the Knight shift is shown in Fig.\ref{fig:K-chi}(a). The observed $K^i$ ($i=1x$, $1y$, $2x$, and $2y$) includes an orbital component $K^i_{\rm orb}$ and a spin one $K^i_{s}$, which are $T$-independent and $T$-dependent, respectively. Here $K^i = K^i_{s} + K^i_{\rm orb} = A_{\rm s}^i \chi^{i} + K^i_{\rm orb}$, where $A^i_{\rm s}$ is the hyperfine coupling constant of the Si nuclei with $4f$-electron spin polarization. $K^i(T)$ is plotted against $\chi^{\rm ab}(T)$ with temperature as an implicit parameter, as shown in Fig. \ref{fig:K-chi}(b). From the slopes of these plots, the hyperfine-coupling constants are estimated as $A^{1x}_{\rm s}=$ 4.6 kOe$/\mu_{\rm B}$ and $A^{1y}_{\rm s}= -$5.7 kOe$/\mu_{\rm B}$ for the Si(1) site, and $A^{2x}_{\rm s}=$ 2.1 kOe$/\mu_{\rm B}$ and $A^{2y}_{\rm s}= -$1.8 kOe$/\mu_{\rm B}$ for the Si(2) site. 
From the intercepts at $\chi^{\rm ab}(T)\to 0$ at a high-$T$ limit in these plots, the orbital shifts are estimated as $K^{1x}_{\rm orb}\approx$ 0\% and $K^{1y}_{\rm orb}\approx$ 0.42\% for the Si(1) site, and $K^{2x}_{\rm orb}\approx$ 0.26\% and $K^{2y}_{\rm orb}\approx$ 0.11\% for the Si(2) site. The large anisotropy in the $A^i_{\rm s}$s indicates that at both Si sites, $K^i_{s}$ originates from the dipolar field generated by the spin density on the Si-$3p$ orbitals polarized by hybridization with the $4f$ electrons at the Ce site. 

\begin{figure}[htbp]
\begin{center}
\includegraphics[width=0.8\linewidth]{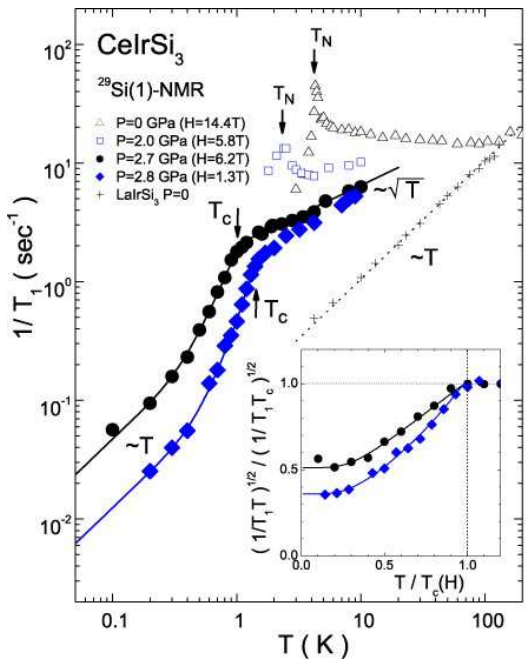}
\end{center}
\caption[]{(Color online) $T$ dependence of $1/T_1$ measured by Si-NMR for CeIrSi$_3$ at $P=0$, 2.0, 2.7 and 2.8 GPa. The solid curves below $T_{\rm c}$ for CeIrSi$_3$ indicate the calculated values obtained by the line-node gap model with $2\Delta_0/k_{\rm B}T_{\rm c}\approx6$ and the RDOS fraction $N_{\rm res}/N_{0}\approx$ 0.37(0.52) in $H=1.3$(6.2) T (see text). The inset shows the plot of $\sqrt{1/T_1T}$ normalized by that at $T_c$ to evaluate $N_{\rm res}/N_{0}$ in low-$T$ limit. }
\label{fig3}
\end{figure}

Next, we present the results of the nuclear-spin lattice relaxation rate $1/T_1$. 
It was measured at a resonance peak corresponding to the $K^{1x}$ in Fig. \ref{fig:spectra}(b), which is separated from the other peaks due to its large Knight shift value. Thus in spite of using the polycrystal sample, the observed $1/T_1$ represents the data for the Si(1) site under a field in the basal plane. As indicated in Fig. \ref{fig3}, a critical divergence in $1/T_1$ was observed at the N\'{e}el temperature $T_{\rm N}\approx$ 4.2 K at $P=0$; this $T_{\rm N}$ is slightly lower than 5 K at $H=0$ \cite{Sugitani}, which may be due to the application of a magnetic field. In the paramagnetic state, $1/T_1$ remains almost constant up to 200 K, suggesting that the $4f$-electron-derived magnetic moments are in a localized regime at $P=0$. By applying the pressure of $P=2$ GPa, the $T_{\rm N}$ determined from a peak in $1/T_1$ decreases to 2.2 K.
At $P=2.7$ GPa, where the AFM order collapses, the $T$ dependence of $1/T_1$ follows a $\sqrt{T}$ behavior in the normal state, in contrast to the $T_1T=constant$ behavior observed in LaIrSi$_3$. When the system is in close proximity to an AFM QCP, the isotropic AFM spin fluctuation model predicts the relation $1/T_1\propto T\sqrt{\chi_{\rm Q}(T)}\propto T/\sqrt{(T + \theta)}$ \cite{Moriya}. Here, the staggered susceptibility $\chi_{\rm Q}(T)$ with the AFM propagation vector ${\bf q}={\bf Q}$ follows a Curie-Weiss law. Note that if $\theta=0$, $1/T_1\propto\sqrt{T}$ would be expected. Thus, $\theta$ is one measure of the proximity of a system to a QCP. In this context, the system at $P=2.7$ GPa, where $1/T_1 \propto \sqrt{T}$, is just in close proximity to the  QCP, revealing $\theta \sim 0$. In the case for $P=2.8$ GPa, the $T$ dependence of $1/T_1$ is well fitted by assuming $\theta=0.5$ K in this model, suggesting the slight deviation from QCP irrespective of a subtle change in $P$. The systematic NMR studies under various $P$ and $H$ will appear in the future.

In the SC state at $P= 2.7$-$2.8$ GPa, $1/T_1$ decreases markedly below $T_{\rm c}(H=6.2$ T$)\approx$1 K and $T_{\rm c}(H=1.3$ T$)\approx$1.4 K; these are in good agreement with the $T_{\rm c}$s determined by the resistivity measurement at magnetic field in the basal plane on a single crystal \cite{Okuda}.  The absence of a coherence peak in $1/T_1$ just below $T_{\rm c}$ and a $T^3$-like dependence upon cooling are evidence of the unconventional nature of superconductivity. 
By assuming a line-node gap with $\Delta=\Delta_0\cos\theta$ and the residual density of states (RDOS) at $E_{\rm F}$ ($N_{\rm res}$), the $1/T_1$ data in the SC state are well fitted by 
\[
\frac{T_1(T_{\rm c})}{T_1(T)}=\frac{2}{k_{\rm B}T}\int\bigl(\frac{N_{\rm s}(E)}{N_0}\bigl)^2 f(E)[1-f(E)]dE,
\]
where $N_{\rm s}(E)=N_0/\sqrt{\Delta^2-E^2}$; $N_0$ is the DOS at $E_{\rm F}$ in the normal state and $f(E)$ is the Fermi distribution function. As shown by the solid curves in Fig. \ref{fig3}, the experimental results are well fitted by assuming $2\Delta_0/k_{\rm B}T_{\rm c}\approx6$ and the RDOS fraction $N_{\rm res}/N_0\approx$ 0.37 (0.52) for $H=1.3 (6.2)$ T. 
It suggests that strong-coupling SC emerges with the line-node gap, which is consistent with the observation of large specific heat-jump at $T_c$\cite{TateiwaCe113}. The $N_{\rm res}/N_{0}$ was determined by the value of $\sqrt{1/T_1T}$ normalized by that at $T_c$ in low-$T$ limit, as shown in the inset of Fig. \ref{fig3}.  It is anticipated to include two contributions arising from the impurity effect and/or from the Volovik effect where the RDOS is induced by a supercurrent in the vortex state \cite{Volovik}. In this case, the contribution of the former is inferred to be approximately 0.3 from the extrapolation to zero field, since the only latter depends on $H$. 

\begin{figure}[htbp]
\begin{center}
\includegraphics[width=0.8\linewidth]{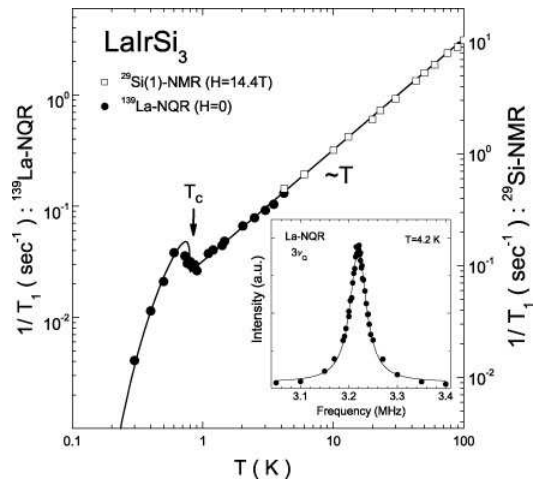}
\end{center}
\caption[]{$T$ dependence of $1/T_1$ measured by Si(1)-NMR and La-NQR for LaIrSi$_3$. The solid curve below $T_{\rm c}$ indicates the calculated values obtained by the isotropic BCS gap model with 2$\Delta_0/k_BT_c=3.1$. The inset is the La-NQR spectrum of 3$\nu_Q$($\pm7/2\Leftrightarrow \pm5/2$). }
\label{fig4}
\end{figure}

The SC state of isostructural compound LaIrSi$_3$ with $T_{\rm c}=$ 0.8 K was investigated by the $^{139}$La-NQR at $H$=0. 
The $^{139}$La-NQR was observed at 3$\nu_Q$($\pm7/2\Leftrightarrow \pm5/2$)= 3.22 MHz with a narrow linewidth less than 40 kHz, ensuring good sample quality, as indicated in the inset of Fig. \ref{fig4}. A distinct coherence peak of $1/T_1$ was observed just below $T_{\rm c}$ as shown in Fig. \ref{fig4}; this was followed by an exponential decrease well below $T_{\rm c}$. These results are well reproduced by the isotropic gap model with $2\Delta_0/k_{\rm B}T_{\rm c}=3.1$ and the parameter of gap broadening $(1/15)\Delta_0$, as indicated by the solid curve in Fig. \ref{fig4}. It is evident that LaIrSi$_3$ is a typical BCS-type $s$-wave superconductor in a weak coupling regime with an isotropic SC gap. From the dHvA measurement, the splitting of the Fermi surface due to the Rashba-type ASOC is estimated to be about 1000 K in LaIrSi$_3$ \cite{Okuda}. 
Even if a large ASOC is present, the typical BCS-type behavior of SC is evident from the NMR experiment. Noting that the NQR intensity is dramatically suppressed below $T_{\rm c}$, even for the powder sample, the penetration depth $\lambda$ is anticipated to be very short, suggesting that LaIrSi$_3$ is a type-I superconductor.

As a result, we have revealed that the SC transition temperature of CeIrSi$_3$ is enhanced up to $T_{\rm c}=$ 1.6 K by the presence of strong AFM spin fluctuations (SFs) around QCP, which is similar to the cases of CeMIn$_5$(M=Co,Rh) with the highest-$T_c$ among the Ce-based HF superconductors \cite{Petrovic,Hegger00}. Here the strong coupling SC state with line node gap has been also realized on the background of the strong AFM SFs as well \cite{IzawaCe115,Yashima}.
It is noteworthy that the $P$-$T$ phase diagram of CeIrSi$_3$ resembles that of CeRhIn$_5$\cite{Hegger00,YashimaPRB2007}.
Both compounds are in the tetragonal structure, however, the AFM SFs of CeMIn$_5$ are characterized by the {\it quasi two-dimensional} (q2D) one evidenced by $1/T_1\sim T/(T+\theta)^{3/4}\sim T^{1/4}(\theta\to 0)$ behavior around the close proximity to the AFM phase \cite{Yashima}. When the heavy quasi-particle bands derived from Ce-$4f$ electron plays an important role, it is likely that the lack of Ce atom at the body center of tetragonal unit cell brings about the q2D AFM SFs in CeMIn$_5$ due to the weaker AFM interlayer coupling than in CeIrSi$_3$ where the Ce atom also occupies the body center of its unit cell. Thus, a possible reason why the highest $T_{\rm c}$ is realized in CeMIn$_5$ is strong AFM SFs enhanced not only by the AFM instability around QCP but also by the lower dimensionality of the electronic structure. 
Nevertheless, the SC in CeIrSi$_3$ can be classified into the group-(b) in Fig.\ref{Yashima-fig1}. 
The most important remaining problem is the influence of the lack of inversion symmetry upon the SC state of CeIrSi$_3$. Actually it has been revealed that the $H_{c2}$ of CeIrSi$_3$ and CeRhSi$_3$ is highly anisotropic, especially, the $H_{c2}$ along the c-axis exceeds more than 30 T, which is extremely higher than the Pauli limiting field\cite{KimuraPRL2007,Okuda}. This kind of behavior has never been observed in CeCoIn$_5$ and CeCu$_2$Si$_2$ evidenced by the spin-singlet Cooper pairing state irrespective of being classified in the same group. 
It suggests that the parity mixing of SC state between even and odd parities may play a crucially role for the noncentrosymmetric superconductor CeIrSi$_3$ through the Rashba-type ASOC. Recent theoretical work has predicted a similar $T$-dependence of $1/T_1$ to that of our result, based on the extended $s + p$ wave Cooper pairing model\cite{Tada}. In order to gain further insight into the possible order-parameter in CeIrSi$_3$, Knight-shift measurement in the SC state using a single crystal is highly desired. 


In conclusion, the SC and magnetic characteristics of the non-centrosymmetric compound CeIrSi$_3$ have been investigated through the $^{29}$Si-NMR measurement under pressure. 
The important finding is that the $1/T_1T$ in CeIrSi$_3$ does not exhibit a coherence peak just below $T_{\rm c}$, and its $T$ dependence is described by the line-node gap model with a large RDOS fraction at the Fermi level.  
We remark that the reason why the maximum $T_{\rm c}$ in CeIrSi$_3$ is relatively high among the Ce-based HF superconductors may be the existence of the strong AFM spin fluctuations, which is similar to the cases of CeMIn$_5$(M=Co,Rh). 
These features resemble those of unconventional superconductors, which are observed in the vicinity of the QCP.
A possible reason for the highest $T_{\rm c}$ realized in CeCoIn$_5$ is strong AFM SFs enhanced not only by the AFM instability around QCP but also by the lower dimensionality of the electronic structure. 

We would like to thank S. Fujimoto, Y. Yanase and M. Yogi for their valuable comments. This work was supported by a Grant-in-Aid for Creative Scientific Research (15GS0213) from the Ministry of Education, Culture, Sports, Science and Technology (MEXT) and the 21st Century COE Program (G18) of the Japan Society of the Promotion of Science (JSPS). 


\end{document}